\documentclass[sn-mathphys-num]{sn-jnl}
\usepackage{geometry}
\usepackage[compat=1.1.0]{tikz-feynman}
\usepackage{graphicx}
\usepackage{slashed}
\usepackage{multirow}%
\usepackage{amsmath,amssymb,amsfonts,bm}%
\usepackage{amsthm}%
\usepackage{mathrsfs}%
\usepackage[title]{appendix}%
\usepackage{xcolor}%
\usepackage{textcomp}%
\usepackage{manyfoot}%
\usepackage{booktabs}%
\usepackage{algorithm}%
\usepackage{algorithmicx}%
\usepackage{algpseudocode}%
\usepackage{listings}%
\usepackage{multicol}%
\usepackage{appendix}%
\usepackage{tikz-3dplot}%
\usetikzlibrary{calc} 
\usetikzlibrary{arrows.meta} 
\usetikzlibrary{decorations.markings} 
\usetikzlibrary{arrows.meta,bending} 
 
\begin{document}

\title {Unitarity implications of $\tilde{F}_{\mu\nu}$ and $F_{\mu\nu}$ models with Lorentz invariance violation in $e^{+}e^{-} \rightarrow \mu^{+}\mu^{-}$}

\author*[1]{\fnm{Alessio} \sur{T. B. Celeste}}\email{alessio.tony@ifpb.edu.br}

\author[2]{\fnm{Adriano} \sur{M. Santos}} 

\affil*[1]{Instituto Federal da Paraíba, Campus Itaporanga, PB-386, Km 2, Itaporanga PB,  58780-000, Brazil} 

\affil[2]{Programa de Pós-graduação em Física, Universidade Federal do Maranhão, Campus Universitário do Bacanga, São Luís MA, 65080-805, Brazil}

\abstract{We studied two different models that included Lorentz Invariance Violation coupling in the scattering processes of $e^+e^- \rightarrow \mu^+\mu^-$. We found that using the model with the dual electromagnetic tensor $\tilde{F}_{\mu\nu}$ resulted in violations of unitarity in both vector and axial scenarios. On the other hand, using the model with nonminimal coupling with $F_{\mu\nu}$ preserved unitarity in both vector and axial cases. As a result, this could have significant implications, given that the nonminimal coupling model with the dual electromagnetic tensor $\tilde{F}_{\mu\nu}$ appeared to be potentially superior to the electromagnetic tensor $F_{\mu\nu}$. Therefore, we believe that these findings could provide a valuable guide for further exploration into the study of CPT and Lorentz breaking phenomena, with significant implications that are certainly nontrivial.}


\maketitle
\section{Introduction}
\begin{multicols}{2}

After the seminal work by Carroll, Field, and Jackiw (CFJ) \cite{Carroll}, the exploration of Lorentz covariance breakdown in Quantum Field Theory (QFT) has been extensively studied, particularly in the context of the Standard Model Extension (SME) developed by Colladay and Kostelecky \cite{Colladay1, Colladay2}. The SME provides a comprehensive framework for investigating Lorentz Invariance Violation (LIV) across high and low-energy physics domains. By adding Lorentz-violating interaction terms to the Standard Model (SM) lagrangian, the SME encompasses various criteria, including power counting renormalizability (\textit{i.e.} with mass dimension $\leq 4$) and gauge invariance ${SU}(3) \times SU(2) \times U(1)$. This investigation covers a wide range of topics including supersymmetry \cite{Berger, Stefan}, non-commutative field theory \cite{Nekrasov, Jabbari, Sean}, and gravitational theories \cite{Moffat}. Another important area of research intersects with the emergence of LIV in string theory \cite{Alan1}, the reintroduction of the concept of preferred reference frames (similar to the old idea of the ether) \cite{Jacobson}, and recent developments in condensed matter physics \cite{Alan2, Urrutia}. It is crucial to consider that the concept of LIV has been seen as a possible solution to persistent mysteries in various areas of physics. One particularly compelling example can be found in the field of cosmic ray physics \cite{Wolfgang}. While LIV may not be the only solution to these complex issues, it often enhances this to make it highly plausible. This suggests that LIV transcends mere peripheral significance in the scope of challenging-to-explore physics, implying substantial implications beyond our current experimental capabilities. Indeed, if alternative explanations fail to adequately address these puzzles, LIV could potentially emerge as an indispensable component of the ultimate theory, surpassing both the SM of particle physics and General Relativity. 

Experimental tests of Lorentz Invariance (LI) stand as paramount endeavors, serving to scrutinize the fundamental symmetries underpinning the fabric of the universe \cite{Mattingly, Bluhm, Mattingly2}. Within this context, understanding LIV becomes imperative. LIV, instead of indicating a loss of covariance, suggests nuanced variances within specific particle sectors, potentially stemming from phenomena such as Quantum Gravity at immensely high energies. Despite Lorentz symmetry enduring all experimental scrutiny within feasible energy scales, the quest for LIV necessitates meticulous precision. Detecting LIV entails remarkable accuracy in low-energy physics experiments and relies on discerning scaling effects in (ultra) high-energy domains. In low-energy realms, tensors are characterized by LIV, while modified dispersion relations categorize it in high-energy contexts. The Planck energy scale $E_P =  \sqrt{\frac{\hbar c^5}{G}} \approx 1.2 \times 10^{19} \, \text{GeV}$ emerges as a pivotal threshold, modulating these relations, yet the substantial energies observed in astrophysical phenomena facilitate detection despite potential suppression. Consequently, experimental inquiries into LI constrain parameters of modified models or LIV functions, demanding particle-specific interpretations. Ensuring that theoretical predictions are validated requires confirmation through empirical evidence. Despite acknowledging the challenges in promptly obtaining such evidence, the efforts outlined in the tables \cite{Alan3} provide an initial glimpse into potential validation methods.

However, it is in the domain of Quantum Electrodynamics (QED) that studies involving LIV become particularly interesting. There are numerous reasons for this; one is that QED is a refined theory that seems to work remarkably well under various scrutinies. Quantities such as the anomalous magnetic moments of the electron and the muon can be measured and experimentally compared with incredible precision. However, efforts from laboratories such as Brookhaven \cite{Bennett} and notably Fermilab \cite{Abi, Aguillard} in recent years have brought anticipation of signals of new physics, given discrepancies between theoretical predictions and increasingly accurate experimental data, forcing a point of attack on one of the most successful theories within the scope of the SM. Especially concerning muons, due to their mass being approximately $4 \times 10^4$ times greater, they are therefore more sensitive to the experimental limits that we have. This sensitivity is essential for probing phenomena beyond the SM. Exploring the high-energy frontier offers a unique avenue to interrogate nature's fundamental laws, motivating endeavors such as muon colliders. The potential to reach multi-TeV energy regimes in a muon collider unlocks opportunities for probing Higgs boson properties. Projects like LEMMA \cite{Antonelli} investigate muon production from $e^+e^-$ annihilation, aiming to exploit the process's threshold for muon pair production. Experimental data near this threshold are scarce, urging precise measurements of production cross sections and muon pair kinematics to validate theoretical predictions. While leading-order QED calculations for $e^+e^- \rightarrow \mu^+\mu^-$ are robust, higher-order radiative effects, particularly due to Coulomb interaction, gain significance near the kinematic threshold. Experimental setups, like those employing 45 GeV positrons on Beryllium or Carbon targets, are devised to study muon pair production in detail \cite{Amapane}. In our previous research \cite{Celeste}, when examining the same electron-positron scattering but producing a photon pair, we found that when the value of $s$ approaches $\infty$, using the dual electromagnetic model, the calculations showed the preservation of unitarity. Given this scenario with the absence of mass, we decided to test this dual model and the electromagnetic tensor model in a massive scenario and analyze what the calculations would reveal.

The structure of this paper unfolds as follows. In Section \ref{part II}, we explore the analysis of the cross-sectional area for $e^+e^- \rightarrow \mu^+\mu^-$  annihilation within the framework of QED. Our investigation entails examining contributions from Feynman diagrams at the tree level. In Section \ref{part III}, we examine a modified version of QED in four dimensions with LIV where the coupling between the photon and the fermions consists of two distinct terms: a minimal coupling and a nonminimal coupling. In Section \ref{part IV}, we present the calculations of the cross sections with the electromagnetic tensor and the dual electromagnetic tensor in the vector and axial scenarios. Finally, in Section \ref{part V}, we discuss and comment on the possible implications of the results obtained.

\section{Muon-pair production by electron-positron scattering} \label{part II}

In the realm of QED, the preeminent influence on determining a cross-sectional area or decay rate typically emanates from the Feynman diagram boasting the scantiest array of interaction vertices, distinguished as the lowest-order diagram. For the annihilation process $e^+e^- \rightarrow \mu^+\mu^-$, a single lowest-order QED diagram presides, elegantly depicted in Figure \hyperref[fig1]{1}. Within this diagram, a duet of QED interaction vertices adorns its architecture, each endowing the matrix element with a factor of $ie\gamma^\mu$, where $e$ is the charge of the electron. Consequently, irrespective of ancillary deliberations, the squared matrix element $|\mathcal{M}|^2$ exhibits a direct proportionality to $e^4$, or equivalently $|\mathcal{M}|^2 \propto \alpha^2$, where $\alpha$ stands as the adimensional fine-structure constant, $\alpha = \frac{1}{4\pi \epsilon_0} \frac{e^2}{\hbar c}$. Every QED vertex contributes a coefficient of $\alpha$ to the expressions dictating cross-sectional areas and decay rates.

\begin{center}
\begin{tikzpicture}
		\begin{feynman}
		  \vertex (a) at (-2.2, -2.5) {$e^{-}$};
		  \vertex (b) at (-2.2, 2.5) {$e^{+}$}; 
            \vertex (c) at (4.7, 2.5) {$\mu^{+}$};		
            \vertex (d) at (4.7, -2.5) {$\mu^{-}$};		
            \vertex (e) at (0, 0);
            \vertex (f) at (2.5, 0); 
            \vertex (g) at (1.2, -0.5) {$\frac{-i g_{\mu\nu}}{(p_1+p_2)^2}$}; 
            \vertex (h) at (-0.45, 0) {$ie\gamma^\mu \,\,\, \bullet$}; 
            \vertex (i) at (2.9, 0) {$\bullet \,\,\, ie\gamma^\nu$}; 
            \vertex (j) at (-1.0, -2.0) {$u(\bm{p}_1,s)$};
            \vertex (k) at (-1.0, 2.0) {$\bar{v}(\bm{p}_2,s')$};
            \vertex (l) at (3.5, 2.0) {$v(\bm{p}_4,r')$};
            \vertex (m) at (3.5, -2.0) {$\bar{u}(\bm{p}_3,r)$};
			\diagram{
                    (a) -- [fermion] (e),
				    (e) -- [photon,momentum=$\gamma$] (f),
                    (b) -- [anti fermion] (e),
                    (f) -- [anti fermion] (c),
                    (f) -- [fermion] (d), 
			};
   \end{feynman}
   \end{tikzpicture}
\end{center}
\begin{center}
    \textbf{Fig. 1} Scattering $e^{+} + e^{-} \rightarrow \mu^{+} + \mu^{-}.$ \label{fig1}
\end{center}

\textcolor{white}{white line}

In Fig. \hyperref[fig1]{1} $u(p)$ represents the electron spinor, $\bar{v}(p)=v^{\dagger}(p)\gamma^{0}$ the adjoint spinor positron, $v(p)$ the anti-muon spinor and $\bar{u}(p)=v^{\dagger}(p)\gamma^{0}$ the adjoint spinor muon. The indices $r, r', s, s' $ represent the spins of particles and antiparticles. In the center-of-mass frame, the momentum vectors are chosen such that $p_{1,2} = (E_{e}, 0, 0, \pm p)$, $p_{3,4} = (E_{\mu}, \pm E_{\mu} \sin\theta, 0, \pm E_{\mu} \cos\theta)$\footnote{$\theta$ is the scattering angle. The azimuth angle $\phi$ cancels out in scattering calculations.}, $E = \sqrt{(pc)^2 + (m_0c^2)^2}$ represents the total energy, and $m_0$ denotes the particle rest mass (\(\sim\) 0.5 \(MeV/c^2\) for the electron and 106 \(MeV/c^2\) for the muon). Using the Feynman rules of QED, we can now write the amplitude for this process: 
\setlength{\abovedisplayskip}{10pt}
\setlength{\belowdisplayskip}{10pt}
\begin{eqnarray}
    i\mathcal{M} &=& \bar{v}^{s'}(p_2)(-ie\gamma^\mu)u^{s}(p_1) \times \nonumber\\
             && \left(\frac{-i g_{\mu\nu}}{s}\right)\bar{u}^{r}(p_3)(-ie\gamma^\nu)v^{r'}(p_4),
             \label{amplitude}
\end{eqnarray}
where $-ie\gamma^\mu$ is vertex factor usual and $\sqrt{s} = p_1 + p_2 = p_3 + p_4$\footnote{$s= E^2_{cm}$, $E_{cm}$ is the total energy in the center-of-mass frame.} is the photon momentum. The amplitude of the Eq. (\ref{amplitude}) depends on the spins of all four particles involved. But in this work, we shall focus on the unpolarized cross-section. Therefore, the unpolarized cross section in the center-of-mass frame can be written in terms of the scattering angle this way\footnote{In this result it was used using the experimental fact that the muon is much heavier than the electron, $m_\mu \approx 207m_e$.}
\begin{eqnarray}
\left(\frac{d \sigma}{d\Omega}\right)_{\text{QED}} = \frac{\alpha^2}{4s}\sqrt{1-\frac{m_{\mu}^2}{E^2}} \times
\,\,\,\,\,\,\,\,\,\,\,\,\,\,\,\,\,\,\,\,\,\,\,\,\,\,\,\,\,\,\,\,\,\,\,\,\,\,\,\nonumber\\
\left[1 + \frac{m_{\mu}^2}{E^2} + \left(1 - \frac{m_{\mu}^2}{E^2}\right)\cos^2\theta \right], \
\end{eqnarray}
where $\alpha \approx 1/137$ is fine-structure constant. In the high-energy assume that the beam energy $E$\footnote{$E = |\bm{p_1}| = |\bm{p_2}| = |\bm{p_3}| = |\bm{p_4}| = E_{c.m.}/2.$} is much greater than either the muon mass $m_{\mu}$, so we found
\begin{eqnarray}
\left(\frac{d \sigma}{d\Omega}\right)_{QED} =\frac{\alpha^2}{4s} \left(1 + \cos^2\theta\right).
\label{QEDchoque}
\end{eqnarray}
The total cross section is obtained by integrating over $d\Omega$, so
\begin{eqnarray}
\left(\frac{d \sigma}{d\Omega}\right)_{total} =\frac{4 \pi \alpha^2}{3s}.
\end{eqnarray}




\section{Models with nonminimal coupling} \label{part III}

The Lorentz violation terms arise as vacuum expectation values of tensors defined at high energy scales. We are proposing an investigation into nonminimal coupling terms in the calculation of cross-section in the process of $e^{+} e^{-} \rightarrow \mu^{+} \mu^{-}$. In this section, we will present four models that lead to the violation of Lorentz and CPT symmetry due to the existence of a background vector field that couples the fermion field to the electromagnetic field. This background four-vector defines a preferred direction in spacetime, thereby violating Lorentz symmetry.

\subsection{Nonminimal coupling with electromagnetic tensor}

\subsubsection{Vectorial nonminimal coupling}
The vectorial nonminimal coupling model is built from a modification in the covariant derivative of the QED:
\begin{eqnarray}
\label{dercovM1}
D^{vect}_{\mu} = \partial_{\mu} + ie A_\mu + i g b^\nu F_{\mu\nu},
\end{eqnarray}
where $g$ is an effective coupling constant (real) with mass dimension $-2$, $b^\nu = (b_0, \vec{b})$ is the CFJ 4-vector responsible for that breaks the Lorentz symmetry and CPT its has dimensions $[b^\nu]=-1$ and assuming that it couples to the electromagnetic field strength $F_{\mu\nu} \equiv \partial_{\mu}A_{\nu} - \partial_{\nu}A_{\mu}$  corresponding to the gauge field $A_\mu$ and whose components are $F_{0i}=-F_{i0}=-E_i$ and $F_{ij}=-F_{ji}=\epsilon_{ijk}B^k$. The commutator\footnote{For the covariant derivative in the minimal model, the commutator is given by $[\partial_\mu - ieA_\mu, \partial_\nu - ieA_\nu]=-ieF_{\mu\nu}$. We assume that the four-vector is constant; therefore, \(b^\nu\) has been taken out of the derivative.} of modified covariant derivatives is $[D^{vect}_{\mu}, D^{vect}_{\nu}]=-iG_{\mu\nu}$ with $G_{\mu\nu}=eF_{\mu\nu}-g\left(b^\nu\partial_\nu F_{\mu\nu} - b^\mu\partial_\mu F_{\nu\mu}\right)$. This CPT-odd modification in the covariant derivative above affects all electron-photon interactions already at tree level. This LIV scenario with the dimension operator 5 has been proposed in Ref. \cite{Bakke} in the context of topological phases. The gauge invariant modified Dirac equation is written in the form:
\begin{eqnarray}
(i \gamma^\mu D^{vect}_{\mu} - m_e)\psi = 0,
\end{eqnarray}
where $m_e$ is electron mass, $\gamma^\mu$ is the Dirac matrix and $\psi$ is a Dirac electron spinor. In the Eq. (\ref{dercovM1}) the term $ie A_\mu$ is called minimum coupling while the term $i g b^\nu F_{\mu\nu}$ of nonminimal coupling both terms are gauge invariant, however, the nonminimal coupling is not renormalizable. Per the analysis in \cite{Gomes}, the background vector 
\begin{eqnarray}
b^\mu \rightarrow - a_{F}{\color{white}.}^{(5)\alpha\mu}_{\,\,\,\,\,\,\,\,\,\,\,\,\,\,\alpha}.  
\end{eqnarray}

The Earth's rotation and its revolution around the Sun imply that the laboratory is a non-inertial reference frame, which introduces a time-dependent relationship in the coefficients for Lorentz violation. Therefore, experimental results concerning these coefficients must be reported within a prescribed inertial reference frame. In standard practice within the literature, this frame is typically taken to be the Sun-centered frame (SCF). Details of SCF are presented in the Appendix. The experimental sensitivity to parity-odd Lorentz violation is therefore suppressed by at least a factor of $10^{-4}$ \cite{Ding, Alan4}. These variations in coefficients occur at harmonics of the annual frequency, and they arise due to the boost $\beta_{\otimes}  \simeq 10^{-4}$ of the Earth in the SCF ($\beta_{\otimes}$ is related to the Earth's average orbital speed) and the boost $\beta_{L} \simeq 10^{-6}$ of the laboratory due to the rotation of the Earth. The transformation between references for the background field would be
\begin{eqnarray}
b^{\mu}_{Lab} = \Lambda^{\mu}_{\nu} b^{\nu}_{Sun},
\end{eqnarray}
the Lorentz matrix $\Lambda^{\mu}_{\nu}$ implementing the transformation from the Sun-centered frame to the laboratory frame is $\Lambda^{0}_{T}=1, \Lambda^{0}_{J} = - \beta^{J}, \Lambda^{j}_{T} = - (\mathcal{R} \cdot \beta)^{j}, \Lambda^{j}_{J} = \mathcal{R}^{j J}$.

The Lagrangian density for this model is given by \
\begin{eqnarray}
\mathcal{L}^{vect} &=& \bar{\psi_i}\left(i\slashed{\partial} - e\slashed{A} - m_i - g b^\mu \gamma^\nu F_{\mu\nu} \right)\psi_i - \nonumber\\
&&\frac{1}{4} F_{\mu\nu}F^{\mu\nu}.
\label{lagvect}
\end{eqnarray}
where $m_i = m_e, m_{\mu}$, $\psi_i = \psi_e, \psi_{\mu}$, $\bar{\psi_i} \equiv \psi_i^\dagger \gamma^0$ is the conjugate spinor. The last term corresponds to the Lagrangian of Maxwell's theory. The presence of nonminimal coupling in lagrangian modifies the vertex involving electron-photon of QED in the following way \cite{Lan-Wu}
\begin{eqnarray}
\label{vertice1}
ie \gamma^\mu \longrightarrow ie \gamma^\mu + \slashed{q}b^{\mu} - (b \cdot q) \gamma^\mu,
\end{eqnarray}
where $q$ represents the four-momentum of the photon. In section \ref{part IV}, the calculations of the cross sections were obtained by replacing this vertex in the scattering amplitude of Eq. (\ref{amplitude}), as well as the modified vertices that will appear in the next models that will be presented below.

\subsubsection{Axial nonminimal coupling}

In this model, the modified Dirac equation is expressed as in the axial non-minimal coupling, as described in \cite{Belich1}:
\begin{eqnarray}
D^{axial}_{\mu} = \partial_{\mu} + i e A_\mu + i g_a \gamma^5 b^\nu F_{\mu\nu},
\end{eqnarray}
with the chiral Dirac matrix $\gamma_5 = \gamma^5 \equiv i\gamma^0\gamma^1\gamma^2\gamma^3$. The Lagrangian density of this model is given by \\
\begin{eqnarray}
\mathcal{L}^{axial} &=& \bar{\psi_i}\left(i\slashed{\partial} - e\slashed{A} - m_i - g_a b^\mu \gamma^\nu \gamma^5  F_{\mu\nu} \right)\psi_i \nonumber\\
&& - \frac{1}{4} F_{\mu\nu}F^{\mu\nu}.
\end{eqnarray}
The five-dimensional non-renormalizable axial term $g_a b^\mu \gamma^\nu \gamma^5 F_{\mu\nu}$ violates {\it C} parity and preserves {\it PT}; therefore, it violates {\it CPT}. Conversely, the vectorial term in Eq. (\ref{lagvect}) preserves {\it C} parity and violates {\it PT}, thus also violating {\it CPT}.
\subsection{Nonminimal coupling with dual electromagnetic tensor}
\subsubsection{Vectorial nonminimal coupling}

In this model, the covariant derivative with vectorial nonminimal coupling is chosen to be \cite{Belich1, Belich2}:
\begin{eqnarray}
\Tilde{D}^{vect}_{\mu} = \partial_{\mu} + ie A_\mu + i \Tilde{g} \Tilde{b}^\nu \Tilde{F}_{\mu\nu},
\end{eqnarray}
where the term CPT-odd $i\Tilde{g}\Tilde{b}^\nu \Tilde{F}_{\mu\nu}$ is gauge invariant and $\Tilde{g}$ is the coupling constant with mass dimension -2, $\Tilde{b}^\mu$ has similar characteristics to $b^\mu$ and $\Tilde{F}^{\mu\nu} = \frac{1}{2} \epsilon^{\mu\nu\alpha\beta}F_{\alpha\beta}$ is the usual dual field-strength electromagnetic tensor with $\epsilon^{0123} = 1$ (Levi-Civita symbol). VSL parameters associated with other nonminimal derivative couplings involving the dual electromagnetic tensor have been limited in Ref. \cite{Charneski} to values $(\Tilde{g} \Tilde{b}^{\mu})\leq 10^{-3} (GeV)^{-1}$. A direct consequence of the non-minimal coupling introduced in $D_{\mu}$ is that scalar particles display a non-trivial magnetic moment. These terms have been used for the perturbative induction of the CFJ see \cite{Petrov, Gazzola}. Per the analysis in \cite{Gomes}, the background vector $b^\mu$ can be written as
\begin{eqnarray}
\Tilde{b}^\mu = \frac{1}{6} \epsilon^{\mu}_{\,\,\nu\alpha\beta}\,a_{F}^{(5)\nu\alpha\beta}.
\end{eqnarray}

The Lorentz-violating Lagrange density for this model is given by \\
\begin{eqnarray}
\Tilde{\mathcal{L}}^{vect} &=& \bar{\psi_i}\left(i\slashed{\partial} - e\slashed{A} - m_i - \Tilde{g} \Tilde{b}^\mu  \gamma^\nu \Tilde{F}_{\mu\nu} \right)\psi_i \nonumber\\
&& - \frac{1}{4} F_{\mu\nu}F^{\mu\nu}.
\label{laggg}
\end{eqnarray}

The vectorial nonminimal coupling can be written
\begin{equation}
\Tilde{g} \Tilde{b}_\nu \Tilde{F}^{\mu\nu} = \Tilde{g} \Tilde{b}_0 \vec{\gamma} \cdot \vec{B} + \Tilde{g} \vec{\Tilde{b}} \cdot \vec{B} \gamma_0 - \Tilde{g} \vec{\gamma} \cdot (\vec{\Tilde{b}} \times \vec{E}),
\end{equation}
where $\Tilde{b}_0$ and $\Vec{\Tilde{b}}$ are the time-like and space-like components of the coefficient $\Tilde{b}_{\mu}$ respectively. We also have that $\gamma^{0}$ is hermitian, $\vec{\gamma}$ is anti-hermitian, and are related to the $\hat{\beta}$ and $\vec{\alpha}$ matrices through: $\gamma^{0} = \hat{\beta}$, $\vec{\gamma} = \hat{\beta} \vec{\alpha}$. We choose the Dirac matrices in the form,
\begin{eqnarray}
    \hat{\beta} = \gamma^{0} = \begin{pmatrix}
1 & 0 \\
0 & -1
\end{pmatrix} \quad \text{and} \quad \vec{\gamma} = \begin{pmatrix}
0 & \vec{\sigma} \\
-\vec{\sigma} & 0
\end{pmatrix}. \
\end{eqnarray}

The presence of nonminimal coupling in Eq. (\ref{laggg}) generates a modification in the vertices as follows (see Ref. \cite{Lan-Wu}),
\begin{eqnarray}
\label{vertice2}
ie \gamma^\mu \rightarrow ie \gamma^\mu - \epsilon^{\mu}_{\,\,\,\alpha\nu\beta}\gamma^{\alpha} \Tilde{b}^{\nu}q^{\beta},
\end{eqnarray}
where $q^{\mu}$ is the photon momentum pointing into the vertex.

\subsubsection{Axial nonminimal coupling}

An axial nonminimal coupling model is obtained by writing the covariant derivative in the form 
\begin{eqnarray}
\label{dercov}
\Tilde{D}^{axial}_{\mu} = \partial_{\mu} + ie A_\mu + i \Tilde{g}_a \gamma^5 \Tilde{b}^\nu \Tilde{F}_{\alpha\beta},
\end{eqnarray}
this model is called torsion-like nonminimal coupling and the Lorentz-violating Lagrange density is written as 
\begin{eqnarray}
\Tilde{\mathcal{L}}^{axial} = &\bar{\psi_i}&\left(i\slashed{\partial} - e\slashed{A} - m_i - \Tilde{g}_a \gamma^5 \Tilde{b}^\nu \gamma^\mu \Tilde{F}_{\alpha\beta} \right)\psi_i \nonumber\\ &-& \frac{1}{4} F_{\mu\nu}F^{\mu\nu}.
\end{eqnarray}

Similar to the previous model, the axial term $\Tilde{g}_a \gamma^5 \Tilde{b}^\nu \gamma^\mu \Tilde{F}_{\alpha\beta}$ violates {\it C} parity and preserves {\it PT}, therefore violates {\it CPT}. Differently the vectorial term in Eq. (\ref{laggg}) preserves {\it C} parity and violates {\it PT}, thus, it also violates {\it CPT}.

\section{Results} \label{part IV}

In this section, we will perform calculations using Feynman's rules. We will consider a modified vertex with a nonminimal CPT-odd coupling term. Specifically, we will analyze the scattering process \( e^+e^- \rightarrow \mu^+\mu^- \) and study its behavior at high energy, where \( p_{1,2}^2 = p_{3,4}^2 = m^2 = 0 \). Throughout our work, we will include scalar products involving the background field up to second order in our results.

\subsection{Vectorial nonminimal coupling with $F_{\mu\nu}$}
As previously mentioned, we derived the scattering amplitude by modifying the QED vertex in Eq. (\ref{amplitude}) with the vertex presented in Eq. (\ref{vertice1}). The resulting cross-section obtained was

{\small
\begin{eqnarray}
\left(\frac{d \sigma}{d \Omega}\right)^{vect} &=& \left(\frac{d \sigma}{d\Omega}\right)_{QED} \,+\,\frac{\alpha^2 g^2}{2s} \left(1 + \cos^2\theta\right) \nonumber\\
&& \,\,\,\,\,\,\,\,\,\,\,\,\,\,\,\,\,\,\,\,\,\,\,\,\,\,\,\,\,\,\,\, \times \left(b \cdot p_1 + b \cdot p_2\right)^2. 
\label{vectelettensor1}
\end{eqnarray}
}
\\
In the case of a purely time-like background, i.e, $b_\mu = (b_0, \bm{0})$, the Eq. (\ref{vectelettensor1}) will be\\
{\small
\begin{eqnarray}\left(\frac{d \sigma}{d \Omega}\right)_{t}^{vect} =&& \left(\frac{d \sigma}{d\Omega}\right)_{QED} \,+\, \frac{\alpha^2 g^2}{2s} \left(1 + \cos^2\theta\right) \nonumber\\ 
&& \,\,\,\,\,\,\,\,\,\,\,\,\,\,\,\,\,\,\,\,\,\,\,\,\,\,\, \times \left(b_0  p^{\,0}_{1} + b_0  p^{\,0}_{2}\right)^2. \
\label{vectelettensor2}
\end{eqnarray}
}
\\
In the case of a purely space-like background, i.e, $b_\mu = (0, \bm{b})$, the Eq. (\ref{vectelettensor1}) will be\footnote{In this result we do not consider terms of the type $1/s^2$ and $1/s^3$.}\\
{\small
\begin{eqnarray}
\left(\frac{d \sigma}{d \Omega}\right)_{s}^{vect} &=& \left(\frac{d \sigma}{d\Omega}\right)_{QED} + \frac{\alpha^2 g^2}{2s} \left(\bm{b} \cdot \bm{p}_1 + \bm{b} \cdot \bm{p}_2\right) \nonumber\\
&&  \,\,\,\,\,\,\,\,\,\,\,\,\,\,\,\,\,\,\,\, \times \left[ \left(\bm{b} \cdot \bm{p}_1 + \bm{b} \cdot \bm{p}_2\right) \left(2 + \cos^2\theta\right) \right. \nonumber\\
&&  \,\,\,\,\,\,\,\,\,\,\,\,\,\,\,\,\,\,\,\, - \left. \left(\bm{b} \cdot \bm{p}_3 + \bm{b} \cdot \bm{p}_4\right)\right]. 
\label{vectelettensor3}
\end{eqnarray}
}
\\
Note in cross sections of the Eqs. (\ref{vectelettensor1}), (\ref{vectelettensor2}) and (\ref{vectelettensor3}) that when $s \to \infty$ the results obtained preserve the unitarity.

\subsection{Axial nonminimal coupling with $F_{\mu\nu}$}
Now, we present the result of the cross-section from the modification of the scattering amplitude Eq. (\ref{amplitude}), we obtained
\begin{eqnarray}
\left(\frac{d \sigma}{d \Omega}\right)^{axial} &&= \left(\frac{d \sigma}{d\Omega}\right)_{QED} + \frac{\alpha^2 g_a^2}{s} \cos\theta \nonumber\\
&&  \,\,\,\,\,\,\,\,\,\,\,\,\,\,\,\,\,\,\,\, \times \left(b \cdot p_1 + b \cdot p_2\right)^2.   
\label{dualelettensor1}
\end{eqnarray}
In the case of a purely time-like background, the Eq. (\ref{dualelettensor1}) will be
\begin{eqnarray}
\left(\frac{d \sigma}{d \Omega}\right)_{t}^{axial} &&= \left(\frac{d \sigma}{d\Omega}\right)_{QED} + \frac{\alpha^2 g_a^2}{s} \cos\theta \nonumber\\
&&  \,\,\,\,\,\,\,\,\,\,\,\,\,\, \times \left(b_0 p^{\,0}_{1} + b_0 p^{\,0}_{2}\right)^2. 
\label{dualelettensor2}
\end{eqnarray}
In the case of a purely space-like background, the Eq. (\ref{dualelettensor1}) will be
%
\begin{eqnarray}
&&\left(\frac{d \sigma}{d \Omega}\right)_{s}^{axial} = \left(\frac{d \sigma}{d\Omega}\right)_{QED} + \, \frac{\alpha^2 g_a^2}{2 s^2} \nonumber \\
&& \,\,\,\,\,\,\,\,\,\,\,\,\,\,\,\,\,\,\,\,  \times \Bigg\{ (\bm{b} \cdot \bm{p}_4)(\bm{b} \cdot \bm{p}_1 + \bm{b} \cdot \bm{p}_2)(p_1^2 - p_2^2) \nonumber\\
&& \,\,\,\,\,\,\,\,\,\,\,\,\,\,\,\,\,\,\,\,  + \, s \left( 3\bm{b} \cdot \bm{p}_1 + \bm{b} \cdot \bm{p}_2 + \bm{b} \cdot \bm{p}_3 + \bm{b} \cdot \bm{p}_4\right) \nonumber\\
&& \,\,\,\,\,\,\,\,\,\,\,\,\,\,\,\,\,\,\,\,  \times (\bm{b} \cdot \bm{p}_1 + \bm{b} \cdot \bm{p}_2)\cos\theta \Bigg\}.
\label{dualelettensor3}
\end{eqnarray}

As in the case of vectorial nonminimal coupling, the cross sections of Eqs. (\ref{dualelettensor1}), (\ref{dualelettensor2}), and (\ref{dualelettensor3}) show that as $s \to \infty$, the obtained results preserve unitarity.

\subsection{Vectorial nonminimal coupling with \textit{\~{F$_{\mu\nu}$}}}
Here, the scattering amplitude was computed by modifying the QED vertex in Eq. (\ref{amplitude}) with the vertex given in Eq. (\ref{vertice2}). The resulting cross-section obtained was:
\begin{eqnarray}
&& \left(\frac{d \sigma}{d \Omega}\right)^{vect} = \left(\frac{d \sigma}{d\Omega}\right)_{QED} + \, \frac{\alpha^2 \Tilde{g}^2}{s}  \nonumber\\
&& \,\,\,\,\,\,\,\,\,\,\,\,\,\,\,\,\,\,\,\,\,\,\,\,\,\,\,\,\,\,\,\,\, \times \left[ \left( \Tilde{b} \cdot p_1 + \Tilde{b} \cdot p_2 \right)^2 -\Tilde{b}^2 s \right] \nonumber\\ 
&& \,\,\,\,\,\,\,\,\,\,\,\,\,\,\,\,\,\,\,\,\,\,\,\,\,\,\,\,\,\,\,\,\, \times \, \left( 1 + \cos^2\theta \right). 
\label{sechoqvl2} 
\end{eqnarray}

In the case of a purely time-like background, the Eq. (\ref{sechoqvl2}) will be
\begin{equation}
\left(\frac{d \sigma}{d \Omega}\right)_{t}^{vect} = \left(\frac{d \sigma}{d\Omega}\right)_{QED} - 3 \alpha^2 \Tilde{b}_0^2 \Tilde{g}^2 \left( 1 + \cos^2\theta \right).
\label{sechoqvl2tempo}
\end{equation}

In the case of a purely space-like background, the Eq. (\ref{sechoqvl2}) will be
\begin{eqnarray}
\left(\frac{d \sigma}{d \Omega}\right)_{s}^{\text{vect}} &=& \left(\frac{d \sigma}{d\Omega}\right)_{QED} + \frac{\alpha^2 \Tilde{g}^2}{s} \nonumber\\
&& \times \left[ \left(\bm{\Tilde{b}} \cdot \bm{p}_1 + \bm{\Tilde{b}} \cdot \bm{p}_2\right)^2 \right. + \left. \bm{\Tilde{b}}^2 s \right] \nonumber\\
&& \times \left( 1 + \cos^2\theta \right).
\label{sechoqvl2espaco}
\end{eqnarray}

Note in cross sections of the Eqs. (\ref{sechoqvl2}), (\ref{sechoqvl2tempo}) and (\ref{sechoqvl2espaco}) that when $s \to \infty$, the obtained results violate unitarity.
\subsection{Axial nonminimal coupling with \textit{\~{F$_{\mu\nu}$}}}
Now we will present the result of the cross-section obtained from the modification of the scattering amplitude given in Eq. (\ref{amplitude}):
\begin{eqnarray}
\left(\frac{d \sigma}{d \Omega}\right)^{axial} &=& \left(\frac{d \sigma}{d\Omega}\right)_{QED} + \frac{2 \alpha^2 \Tilde{g}_a^2}{s} \cos\theta \times \nonumber\\
&&\left[ \left( \Tilde{b} \cdot p_1 + \Tilde{b} \cdot p_2\right)^2 \right.  - \left. s \Tilde{b}^2 \right].
\label{sechoqvl3}
\end{eqnarray}
In the case of a purely time-like background the Eq. (\ref{sechoqvl3}) will be
\begin{eqnarray}
\left(\frac{d \sigma}{d \Omega}\right)_{t}^{axial} = \left(\frac{d \sigma}{d\Omega}\right)_{QED} - 6 \alpha^2 \Tilde{b}_0^2 \Tilde{g}_a^2 \cos\theta.\,\,\,
\label{sechoqvl3tempo}
\end{eqnarray}
In the case of a purely space-like background the Eq. (\ref{sechoqvl3}) will be
\begin{eqnarray}
\left(\frac{d \sigma}{d \Omega}\right)_{s}^{\text{axial}} &=& \left(\frac{d \sigma}{d\Omega}\right)_{QED} + \frac{2 \alpha^2 \Tilde{g}_a^2}{s} \cos\theta \times \nonumber\\
&&\left[ \left( \bm{\Tilde{b}} \cdot \bm{p}_1 + \bm{\Tilde{b}} \cdot \bm{p}_2\right)^2 + s \bm{\Tilde{b}}^2\right]. \
\label{sechoqvl3espaco}
\end{eqnarray}

As with vectorial nonminimal coupling, the cross sections in Eqs. (\ref{sechoqvl3}), (\ref{sechoqvl3tempo}), and (\ref{sechoqvl3espaco}) violate unitarity as $s \to \infty$.

It is possible to find an upper bound for the LIV parameters. The QED cutoff parameters are defined by \cite{Bender}
\begin{eqnarray}
\left| \frac{\sigma}{\sigma_{QED}} -1 \right| = \left( \frac{2s}{\Lambda_{\pm}^{2}} \right),
\end{eqnarray}
where $\Lambda_{\pm}$ is a small parameter representing possible experimental departures from the theoretical prediction. Assigning the values for CM energies  $\sqrt{s} = 29$ GeV and QED cutoff parameters $\Lambda_{\pm} = 172$ GeV in $e^{+}e^{-} \rightarrow \mu^{+}\mu^{-}$ \cite{Derrick}, we find from Eqs. (\ref{QEDchoque}) and (\ref{vectelettensor2})
\begin{eqnarray}
(g b_0) \leq 6 \times 10^{-3} (GeV)^{-1}.
\end{eqnarray}
Dimension-five terms with $F_{\mu\nu}$ couplings in \cite{Sherrill} using the term $- \frac{1}{2} (a_{F}^{(5)})^{\mu\alpha\beta}_{AB} F_{\alpha\beta} \bar{\psi}_{A}\gamma_{\mu} \psi_{B}$ indicates a range $10^{-13} - 10^{-9} (GeV)^{-1}$ for the LIV coefficient.

The Fig. (\ref{Figure2}) illustrates the representation of the angular dependence observed in the differential cross-section. This depiction specifically pertains to $e^{+}e^{-} \rightarrow \mu^{+}\mu^{-}$ scattering under the influence of time-like Lorentz violation and vectorial nonminimal coupling
\setcounter{figure}{1}
\begin{figure} [H]
    \centering
    \includegraphics[width=1\linewidth]{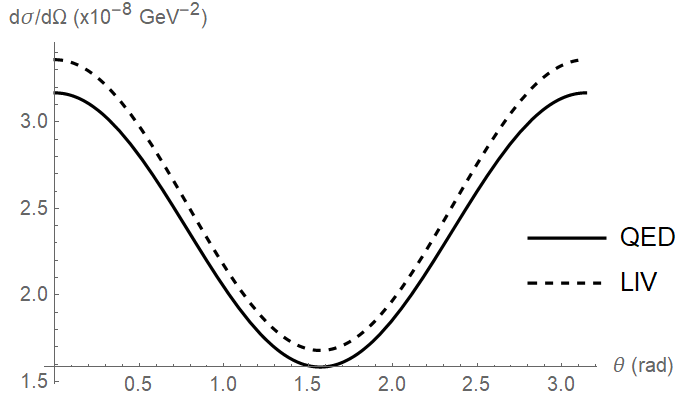}
    \caption{The differential cross sections for the usual QED (solid line) and the timelike LIV contribution (dashed line).}
    \label{Figure2}
\end{figure}

\section{Final remarks} \label{part V}
In our research, we have studied the violation of Lorentz symmetry by examining the production of muon pairs through electron-positron scattering within the framework of Extended Quantum Electrodynamics. We tested two models involving nonminimal coupling in the scattering process \( e^+e^- \rightarrow \mu^+\mu^- \), and calculated the cross-section. Our results show that when using the dual electromagnetic tensor, both the vector and axial scenarios violated unitarity. On the other hand, using the model with the preserved electromagnetic tensor maintained unitarity in both scenarios, indicating its potential superiority over the former model. Furthermore, radiative corrections calculations \cite{Lan-Wu} in massless QED have demonstrated the renormalizability of the model with radiative corrections at 1-loop, a property not observed in the model involving the dual field. This further strengthens our confidence in the model with \( F_{\mu\nu} \) coupling. In the future, we plan to explore the potential phenomenological implications of this model in various contexts. Ultimately, we believe that these findings may provide a valuable roadmap for further investigation into the study of CPT and Lorentz breaking.
\

\begin{appendices}
\renewcommand{\theequation}{A.\arabic{equation}}  
\section*{Appendix A: Sun-centered celestial equatorial frame} \label{part VI}

The canonical inertial frame adopted in the literature is the Sun-centered celestial equatorial frame (SCF), which has the origin of its time coordinate $T$ defined as the 2000 vernal equinox. The Cartesian coordinates in this frame are specified as having the Z axis aligned along the rotation axis of the Earth and the $X$ axis pointing from the Earth to the Sun, with the $Y$ axis completing a right-handed coordinate system, see Figure \hyperref[figA1]{A.1}. The SCF is well suited as a standard frame because it is essentially inertial during typical experimental time scales and because its axes are conveniently chosen for laboratory studies.

The relationship $x^{j} = R^{jJ}x^{J}$ between the coordinates $x^j$ in the laboratory frame ($j = x, y, z = 1, 2, 3$ denotes an index in the laboratory frame) and the coordinates $x^J$ in the SCF ($J = X, Y , Z$ denotes an index in the SCF) is then given by the local sidereal time-dependent rotation matrix $R^{jJ}$. 
\begin{equation}
\scalebox{0.83}{$ R^{jJ} = \begin{pmatrix}
\cos \chi \cos \omega_{\otimes} T_{\otimes} & \cos \chi \sin \omega_{\otimes} T_{\otimes} & - \sin \chi \\
- \sin \omega_{\otimes} T_{\otimes} & \cos \omega_{\otimes} T_{\otimes} & 0 \\
\sin \chi \cos \omega_{\otimes} T_{\otimes} & \sin \chi \sin \omega_{\otimes} T_{\otimes} & 0
\end{pmatrix}
$}.
\end{equation}
In this matrix, $\chi$ is the colatitude of the laboratory (angle between the beam direction and the $Z$ axis), $\omega_{\otimes} \simeq 2\pi/(23h56 min)$ is the Earth’s sidereal angular frequency, $T_{\otimes}$ is measured in the SCF from one of the times when the y- and Y -axes coincide, to be chosen conveniently for each experiment. The matrix generates the harmonic time dependences of the coefficients for Lorentz violation observed in the laboratory frame. In addition to its sidereal rotation, the Earth’s revolution about the Sun induces further time variations in the coefficients for Lorentz violation in the laboratory and apparatus frames. Implementing this transformation on the cross-section generates a lengthy-expression containing up to fourth harmonics of the sidereal frequency \cite{Li}. 
The velocity 3-vector of the laboratory in the SCF is
\begin{equation}
\scalebox{0.82}{$  \boldsymbol{\Vec{\beta}} = \boldsymbol{\beta}_\oplus
\begin{pmatrix}
\sin \Omega_\oplus T \\
- \cos \eta \cos \Omega_\oplus T \\
- \sin \eta \cos \Omega_\oplus T
\end{pmatrix}
+ \boldsymbol{\beta}_L
\begin{pmatrix}
- \sin \omega_\oplus T_\oplus \\
\cos \omega_\oplus T_\oplus \\
0
\end{pmatrix}
$},
\end{equation}
where the boost \( \boldsymbol{\beta}_\oplus = R_\oplus \Omega_\oplus \) is the speed of the Earth’s orbital motion (\( R_\oplus \) is the mean orbital radius and \( \Omega_\oplus \) the angular frequency of the Earth’s orbital motion). The quantity \( \eta \approx 23.4^\circ \) is the angle between the XY celestial equatorial plane and the Earth’s orbital plane. The boost \( \boldsymbol{\beta}_L = r_\oplus \omega_\oplus \sin \chi \lesssim 1.5 \times 10^{-6} \) is the speed of the laboratory due to the rotation of the Earth with \( r_\oplus \) being the radius of the Earth.
\end{appendices}

\colorlet{myred}{red!80!black}
\colorlet{myblue}{blue!85!black}
\colorlet{mygreen}{green!75!black}
\colorlet{suncol}{yellow!90!orange}
\colorlet{watercol}{blue!50!cyan!90!black}

\tikzset{
  >=latex, 
  axis/.style={thick,line cap=round},
  whiteline/.style={white,line width=2,line cap=round},
  orbit/.style={#1,thick,rotate around x=\Iorb},
  filled orbit/.style={orbit=#1,fill=blue!30,fill opacity=0.8},
  body/.style={draw=none,ball color=#1,postaction={%
    fill=#1,fill opacity=0.6,draw=#1!40!black,line width=0.05}},
  smallarr/.style={-{Latex[length=3,width=2,flex'=1]}},
  smallerarr/.style={-{Latex[length=2,width=1.5,flex'=1]},
                     line cap=round,line width=0.3},
  orbit/.default={myblue}
}

\def\clipSun#1{ 
  \clip[rotate around x=#1] 
    (#1:\Rsun*\ptunit) arc(#1:180+#1:\Rsun*\ptunit)
    (0,0) circle (\Rsun);
}
\def\drawEarth{
  \begin{scope}[rotate around x=\Iorb]
    \coordinate (E) at (\angE:\Rorb);
    \coordinate (Reta) at (90:\Rorb); 
  \end{scope}
  \drawEarthAxis
  \draw[body=watercol] (E) circle (\Rearth*\ptunit);
  \draw[orbit,line cap=round] (0:\Rorb) arc(0:\angE-5:\Rorb);
  \node[scale=1,anchor=184,inner sep=6pt] at (E) {Earth};
}
\def\drawEarthAxis{
  \def\Laxis{0.5} 
  \def\angarr{135} 
  \begin{scope}[rotate=-12]
  \coordinate (EA) at ($(E)+(0,0,0.40*\Laxis)$);
    \draw[thin,line cap=round] 
      (E)++(0,0,-\Laxis/2) --++ (0,0,\Laxis);
    \draw[white,line width=0.6]
      (EA)++(\angarr:\Rearth*\ptunit)
      arc(\angarr:400:{\Rearth*\ptunit} and {0.05*\ptunit});
    \draw[smallerarr]
      (EA)++(\angarr:\Rearth*\ptunit)
      arc(\angarr:440:{\Rearth*\ptunit} and {0.05*\ptunit});
  \end{scope}
}
\def\drawInclination{
  \def\Rarr{1.5} 
  \pgfmathsetmacro\y{\Rorb*cos(\Iorb)}
  \pgfmathsetmacro\z{\Rorb*sin(\Iorb)}
  \coordinate (Reta') at (0,{\Rsun*cos(\Iorb)},{\Rsun*sin(\Iorb)});
  \draw[dashed,red] (Reta') -- (Reta) -- (0,\y,0);
  \draw[smallarr,red,rotate around y=90,rotate around z=90]
    (\Rarr,0) arc(0:\Iorb:\Rarr) node[pos=0.4,right=-1pt,scale=0.9] {$\eta$};
}

\tdplotsetmaincoords{70}{118} 

\raggedbottom

\begin{flushleft}

\begin{tikzpicture}[tdplot_main_coords]
  
  \def\xmin{-3.6}     
  \def\xmax{4.0}      
  \def\zmin{-1.8}     
  \def\zmax{2.0}      
  \def\ptunit{28.4pt} 
  \def\Rsun{0.25}     
  \def\Rearth{0.11}   
  \def\Rorb{3.0}      
  \def\Iorb{13}       
  \def\angE{127}      
  
  \draw[orbit] (0:\Rorb) arc(0:-180:\Rorb);
  
  \draw[thick,line cap=round] (0,0,0) -- (\xmin,0,0);
  \draw[whiteline] (0,0,0) -- (0,\xmin,0); 
  \draw[thick,line cap=round] (0,0,0) -- (0,\xmin,0);
  \draw[thick,line cap=round] (0,0,0) -- (0,0,\zmin);
  
  \draw[body=suncol] (0,0) circle (\Rsun*\ptunit);
  \draw[whiteline] (-80:\Rorb) arc(-80:-30:\Rorb); 
  \draw[whiteline] (20:\Rorb) arc(20:50:\Rorb); 
  \draw[black] (0,0) circle (\Rorb);
  \draw[whiteline] (0,0,1.4*\Rsun) -- (0,0,\zmax); 
  
  \draw[axis,->]
    (\Rsun,0,0) -- (\xmax,0,0) node[anchor=20] {$\hat{X}$};
  \draw[axis,->]
    (0,\Rsun,0) -- (0,\xmax,0) node[anchor=-160] {$\hat{Y}$};
  \draw[axis,->]
    (0,0,\Rsun) -- (0,0,\zmax) node[anchor=-90] {$\hat{Z}$};

  \draw[whiteline,rotate around x=\Iorb] 
    (20:\Rorb) arc(20:\angE-5:\Rorb);
  \draw[orbit,postaction={decorate},decoration={%
    markings,mark=at position 0.89 with {\arrow{latex}}}]
    (0:\Rorb) arc(0:180:\Rorb);
  
  \drawEarth
   \node[scale=1,anchor=-39,inner sep=7pt] at (0,0) {Sun};
  \drawInclination
  
   \node at (-3.0, -3.0) {$T=0$};

\end{tikzpicture}

\end{flushleft}

\begin{center}
     \textbf{Fig. A.1} Standard Sun-centered inertial reference frame. \label{figA1} \\\
\end{center} 

 \textbf{Acknowledgments}:
We express our gratitude to Prof. J. A. Helayël-Neto for the helpful discussions and suggestions. AMS extends its sincere appreciation to the CAPES (Brazilian Research Agency) for the financial support granted through the Graduate Scholarship. Lastly, the authors would like to express their gratitude to the anonymous referee for their precise and thoughtful comments, which contributed to improving the manuscript.

\end{multicols}
\end{document}